\documentclass[
amsmath, floats,
floatfix,showkeys,
12pt,
nofootinbib,
tightenlines,
showpacs]{revtex4}
\usepackage{hyperref, subfigure, fancyhdr, amsmath, amssymb, url, graphicx, enumerate, verbatim, listings, dsfont, color, bm, slashed, array, mathtools}

\newcommand{\beq}{\begin{equation}}
\newcommand{\eeq}{\end{equation}}

\renewcommand{\v}[1]{\ensuremath{\mathbf{#1}}} 
 
\newcommand{\mbraket}[3]{\left< #1 \vphantom{#2#3} \right|
 #2 \left| #3 \vphantom{#1#2} \right>} 
\let\bar=\smallbar 
\newcommand{\bar}[1]{\overline{#1}} 
\newcommand{\fs}[1]{\slashed{#1}} 

\usepackage{epstopdf}

\begin{document}

\title{\bf Charge Symmetry Breaking and Parity Violating Electron-Proton Scattering\\\hskip12cm NT@UW-14-02}

\author{ Michael Wagman and Gerald A. Miller }

\affiliation{Department of Physics, University of Washington, Seattle, WA 98195-1560}

\date{\today} 

\begin{abstract}
  {Charge symmetry breaking contributions to the proton's neutral weak form factors must be understood in order for future measurements of parity violating electron-proton scattering to be definitively interpreted as evidence of proton strangeness. We calculate these charge symmetry breaking form factor contributions using chiral perturbation theory with resonance saturation estimates for unknown low-energy constants. The uncertainty of the leading-order resonance prediction is reduced by incorporating nuclear physics constraints. Higher-order contributions are investigated through phenomenological vertex form factors. We predict that charge symmetry breaking form factor contributions are an order of magnitude larger than expected from na{\"i}ve dimensional analysis but are still an order of magnitude smaller than current experimental bounds on proton strangeness. This is consistent with previous calculations using chiral perturbation theory with resonance saturation.}

\end{abstract}

\maketitle \noindent

\section{Introduction}\label{intro}

The proton's neutral weak form factors can be determined from measurements of parity violating electron-proton scattering. Assuming charge symmetry, that is invariance under an isospin rotation exchanging $u$ and $d$ quarks, these neutral weak form factors can be identified with a linear combination of nucleon electromagnetic and strangeness form factors. This allows measurements of parity violating electron-proton scattering to directly probe strangeness in the nucleon. Present scattering measurements do not provide conclusive evidence for nucleon strangeness, but more precise measurements are possible~\cite{Armstrong:2012bi}.

Charge symmetry is slightly broken in nature by the $u$ and $d$ quark mass difference and by electromagnetic effects, for reviews see~\cite{Henley:1979ig,Miller:1990iz,Miller:1994zh,Miller:2006tv}. When charge symmetry breaking (CSB) effects are included, there are additional contributions to the proton's neutral weak form factors~\cite{Dmitrasinovic:1995jt,Miller:1997ya,Lewis:1998iu, Kubis:2006cy}. CSB effects are typically small, for example the proton-neutron mass difference is one part in a thousand, but unexpectedly large CSB contributions to the proton's neutral weak form factors could be falsely interpreted as signals of proton strangeness in future experiments. It is important to understand whether uncertainty about CSB effects limits our ability to interpret measurements of new contributions to the proton's neutral weak form factors as signals of proton strangeness.

Non-relativistic quark models predict that CSB form factor contributions vanish at zero momentum transfer and can be safely ignored at low momentum transfer~\cite{Dmitrasinovic:1995jt, Miller:1997ya}. The more general $SU(6)$ quark models used in Ref.~\cite{Miller:1997ya} included CSB effects due to quark kinetic energy differences, one-gluon-exchange operators, and one-photon-exchange operators. 

\begin{figure}
  \includegraphics[width=.8\textwidth]{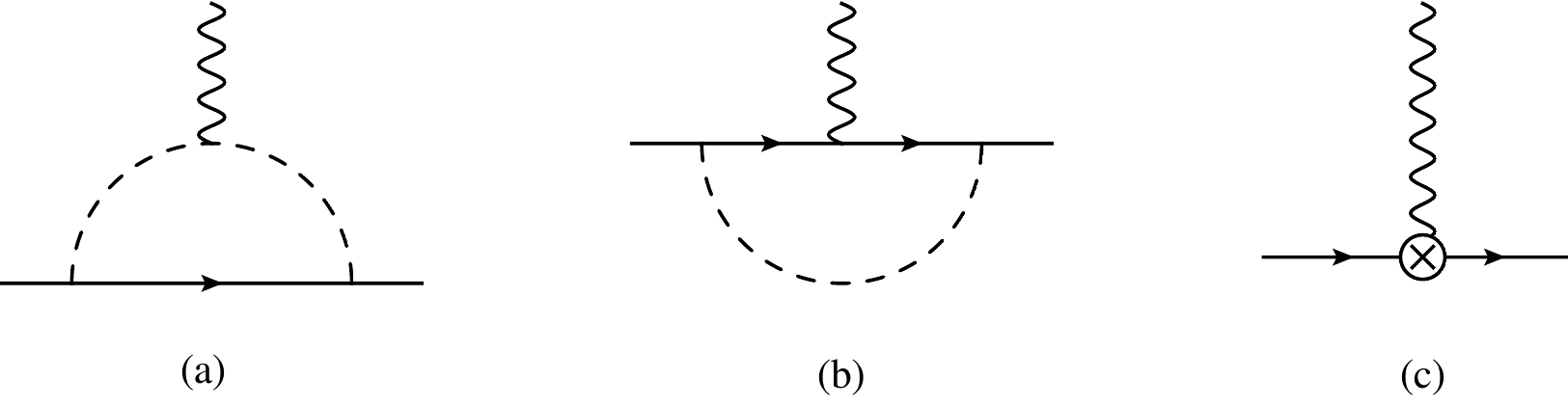}
	\caption{The leading CSB contributions to the proton's neutral weak form factors in chiral perturbation theory. CSB effects arise in the pion loop diagrams (a) and (b) from the proton-neutron mass difference. The crossed circle in diagram (c) represents a CSB nucleon-photon interaction  
	arising from short distance interactions that contributes at the same expansion order in chiral perturbation theory. Wave function renormalization also gives a CSB contribution not shown.} 
	\label{diagrams}
\end{figure}

Additional CSB form factor contributions involving the pion cloud of the nucleon arise in chiral perturbation theory ($\chi$PT). Lewis \& Mobed considered one-pion-exchange contributions in heavy baryon chiral perturbation theory (HB$\chi$PT), shown diagrammatically in Fig.~\ref{diagrams}, where CSB effects result from the proton-neutron mass difference~\cite{Lewis:1998iu}. An unambiguous HB$\chi$PT prediction for the CSB contribution to the neutral weak magnetic form factor could not be made because a CSB nucleon-photon contact interaction unconstrained by symmetry or experiment contributes at leading-order (LO) in chiral power counting.

With experimental measurements or first principles calculations of the size of this contact term, $\chi$PT would predict the CSB contributions to the neutral weak magnetic moment, charge radius, and magnetic radius up to parametrically suppressed error. Without such measurements, $\chi$PT calculations require a model estimate for the strength of the CSB nucleon-photon contact interaction. Kubis \& Lewis (KL)~\cite{Kubis:2006cy} used the resonance saturation technique of Ecker \emph{et al.}~\cite{Ecker:1988te} to estimate the contact term in a resonance exchange model where CSB is driven by $\rho-\omega$ mixing. Combining this estimate with calculations in HB$\chi$PT and infrared regularized baryon chiral perturbation theory, KL predicted a CSB magnetic moment contribution of $0.025\pm 0.02$ including resonance parameter uncertainty~\cite{Kubis:2006cy}. This effect is an order of magnitude smaller than current experimental uncertainties in nucleon strangeness measurements, but it is larger than predictions based on non-relativistic quarks models or na{\"i}ve dimensional analysis.\footnote{The heavy scales present in $\chi$PT are on the order of the nucleon mass. One expects the CSB magnetic moment contribution arising from the proton-neutron mass difference to be suppressed relative to the order one isospin conserving magnetic moment by the $1/1000$ ratio of these scales.}

The  theoretical uncertainty (although small)  has caused  experimentalists to stop their efforts to discover strangeness in nucleons through elastic electromagnetic form factors. For example
Ref.~\cite{Acha:2006my} states ``Theoretical uncertainties especially regarding the assumption of charge symmetry [24], preclude significant improvement to the measurements reported here."
(Ref. [24] of  \cite{Acha:2006my} is our  ~\cite{Kubis:2006cy}.) Similar remarks are made in ~\cite{Paschke:2011zz}. However, Ref.~\cite{Wang:1900ta} states that  the charge symmetry effect ``estimated in the calculation of Kubis and Lewis [53] is an exception" to the general experience that charge symmetry  breaking effects being very small and that 
``implications of this work [53] for other examples of charge symmetry violation have not yet been worked out."  The statements of Ref.~\cite{Wang:1900ta} originate from the strong vector-meson nucleon coupling constants that Kubis \& Lewis employ in their resonance saturation procedure. These coupling constants are a focus of the present work. We also note that  Ref.~\cite{GonzalezJimenez:2011fq} simply states ``isospin violations ... are expected to be very small." So there seems to be a divergence of opinion regarding the importance  of the charge symmetry breaking effects. Given the  large interest in the strangeness content of the nucleon, it is of considerable relevance  to re-examine the charge symmetry breaking effects, and we    do that here.

There were two findings in the work of KL. The first is that the pion loop contribution is relatively large, and the second is the importance of the effects of $\rho-\omega$ mixing. 
In the present work we revisit CSB contributions to the proton's neutral weak form factors using a relativistic form of $\chi$PT, resonance saturation, and we also impose well-known constraints arising
from mass differences between mirror  nuclei that are caused  by CSB effects~\cite{Miller:1990iz}. Sec.~\ref{chiPT} presents a LO calculation of the CSB form factors in relativistic $\chi$PT. Higher order effects are investigated in this framework through phenomenological vertex form factors discussed in Sec.~\ref{vff}. Estimation of the unconstrained counterterm through resonance saturation is discussed in Sec.~\ref{vm}. Sec.~\ref{results} argues that the $\omega$-nucleon coupling constant $g_\omega \sim 42$ used by KL is incompatible with experimental constraints on this coupling constant~\cite{Dumbrajs:1983jd,Ericson:1988gk}
as well as  the $^3$He-$^3$H binding energy difference. Numerical results incorporating these constraints are presented. Our results for the CSB form factors are summarized and relativistic and heavy baryon results are compared in Sec.~\ref{sum}.

\section{Formalism}\label{formalism}
\newcommand{\eqn}[1]{\label{#1}}
\newcommand{\eq}[1]{Eq.~(\ref{#1})}
Without assuming charge symmetry, the proton's neutral weak form factors $G^{p,Z}$ are given by~\cite{Armstrong:2012bi}
\begin{equation}
  G^{p,Z}(Q^2) = \bigl(1-4\sin^2\theta_W \bigr) G^{p}(Q^2) - G^{n}(Q^2) - G^{s}(Q^2) - G^{CSB}(Q^2),
  \label{GpZ}.
\end{equation}
where $q_\mu=p'_\mu-p_\mu$ is the momentum transferred to the nucleon and $Q^2=-q^2$. $G$ represents electric or magnetic Sach's form factors for a particular matrix element, denoted by a superscript. The electric and Sach's form factors for a given matrix element are defined in terms of the corresponding Dirac and Pauli form factors by
\begin{eqnarray}
	\label{Sachs}
	G_E(Q^2) &=& F_1(Q^2) - \frac{Q^2}{4m_N^2} F_2(Q^2),\\\nonumber
	G_M(Q^2) &=& F_1(Q^2) + F_2(Q^2).
\end{eqnarray}
$G^p$ and $G^n$ denote form factors for matrix elements of the light quark electromagnetic current $\frac{2}{3}\bar{u}\gamma^\mu u - \frac{1}{3}\bar{d}\gamma^\mu d$ in proton and neutron states respectively. $G^s$ denotes form factors for matrix elements of the strange quark electromagnetic current $-\frac{1}{3}\bar{s}\gamma^\mu s$ in either nucleon state (the difference between proton and neutron strangeness is ignored). $G^{CSB}$ denotes the CSB from factor contribution and is defined by Eq.~\eqref{Sachs} in terms of the Dirac and Pauli form factors
\begin{eqnarray}
	\label{defFvs}
  \bar{u}(p^\prime)\left[ \gamma^\mu F_1^{CSB}(Q^2) + \frac{i\sigma^{\mu\nu}q_\nu}{2m_N}F_2^{CSB}(Q^2) \right]u(p)&=&  \mbraket{p(p^\prime)}{-\frac{1}{3}\bar{u}\gamma^\mu u + \frac{2}{3}\bar{d}\gamma^\mu d}{p(p)}\\\nonumber\\\nonumber
	&&\hspace{10pt} + \mbraket{n(p^\prime)}{-\frac{2}{3}\bar{u}\gamma^\mu u + \frac{1}{3}\bar{d}\gamma^\mu d}{n(p)}.
\end{eqnarray}
If charge symmetry holds, the right hand side of Eq.~\eqref{defFvs} vanishes. For comparison note that
$G_{E,M}^{CSB} = G_{E,M}^{u,d}$ in the notation of KL. CSB arises from neutron-proton mass difference effects on $F_{1,2}$. The value of $m_N^2$ used in \eq{Sachs} is taken as fixed and  does not cause any CSB.

It is worthwhile to defiance the leading moments of the CSB form factors. These are given by
\begin{eqnarray}
G^{CSB}_{E,M}(Q^2)=G^{CSB}_{E,M}(0)-\rho_{E,M}^{CSB} Q^2 + O(Q^4).
\end{eqnarray}

\subsection{Baryon Chiral Perturbation Theory}\label{chiPT}

Form factors and other hadronic observables can be systematically described by effective field theories (EFTs) such as HB$\chi$PT~\cite{Weinberg:1978kz,Gasser:1983yg,Gasser:1987rb,Jenkins:1990jv}. In HB$\chi$PT the infinite set of pion and nucleon interactions consistent with the symmetries of QCD are organized according to a power counting scheme where for example pion momenta and quark masses are treated as light energy scales, for reviews see Refs.~\cite{Kaplan:2005es,Scherer:2002tk,Bernard:1995dp}. Contributions to observables at each expansion order generally include tree-level contributions from operators of that order and loop-level contributions involving lower order operators. Each operator is parametrized by a low-energy constant (LEC) that can in principle be calculated from QCD. While efforts to compute LECs from lattice QCD are promising, most LECs are still determined phenomenologically by matching calculated observables to experimental data. Without sufficient data these LECs must be estimated through techniques such as resonance saturation or na{\"i}ve dimensional analysis.

In relativistic baryon chiral perturbation theory (RB$\chi$PT), loop contributions to observables can include pieces that violate HB$\chi$PT power counting.\footnote{An approach called infrared regularization consistently reabsorbs these pieces into the LECs of the theory in order to give a manifestly relativistic theory the power counting scheme of HB$\chi$PT~\cite{Becher:1999he}. By RB$\chi$PT we refer to theories that do not remove these power counting violating terms. Infrared regularized loop contributions match HB$\chi$PT loop contributions up to higher-order terms by construction and we will therefore not distinguish between infrared regularized and HB$\chi$PT results.}  RB$\chi$PT and HB$\chi$PT give identical predictions for physical observables but may have different divisions between loop and counterterm contributions to observables. Comparing the loop contributions to the CSB form factors in RB$\chi$PT and HB$\chi$PT probes the sensitivity of this loop/counterterm division to changes in the ultraviolet treatment of the theory that may not be captured by model counterterm estimates. The loop contribution to the CSB magnetic moment is renormalization scale dependent in HB$\chi$PT but not in RB$\chi$PT, and so this comparison probes the sensitivity of the loop/counterterm division at the particular renormalization scale chosen according to resonance saturation prescriptions.

By HB$\chi$PT power counting arguments clearly reviewed in Ref.~\cite{Kubis:2006cy}, the LO contributions to the CSB form factors come from the diagrams of Fig.~\ref{diagrams}. In HB$\chi$PT the only next-to-leading-order (NLO) contributions to $G_M^{CSB}$ come from Fig.~\ref{diagrams}(b) with a nucleon electromagnetic tensor coupling. We have computed the NLO tensor contributions in RB$\chi$PT and found them to be numerically subleading ($\sim 10\%$ of LO results). However, NLO power counting can be ambiguous in RB$\chi$PT and in particular there are NLO contributions from two loop diagrams that vanish in HB$\chi$PT but might have power counting violating contributions in RB$\chi$PT. These contributions could lead to renormalization of the unconstrained nucleon-photon contact interaction at NLO in RB$\chi$PT. Clearer estimates for the size of higher-order RB$\chi$PT corrections will be discussed in Sec.~\ref{vff} and in this section we will only describe a RB$\chi$PT calculation at LO.

The chiral Lagrangian pieces required for a calculation of the CSB form factors are given in a compact relativistic notation in Ref.~\cite{Kubis:2006cy} and in heavy baryon form in Ref.~\cite{Lewis:1998iu}. For quick reference in more pedestrian relativistic notation, the nucleon Lagrangian terms needed for a LO calculation of the CSB form factors are
\begin{eqnarray}
  \mathcal{L}_{N\pi\gamma} &=& \bar{N}\left[ i\fs{\partial} -Q\fs{A} - \left( m_N - \frac{\Delta m_N}{2}\tau_3 \right) - \frac{g_A}{2f_\pi}\partial_\mu \pi^a \gamma^\mu \gamma_5 \tau^a\right.\label{LNpigamma}
\\\nonumber
  &&\hspace{25pt} \left.+ \frac{e\sigma^{\mu\nu}}{8m_N}F_{\mu\nu}(\kappa^{\fs{v}} + \kappa^{\fs{s}}\tau^3)  \right]N.
	\end{eqnarray}
In this $N = (p, n)^T$ is an isospinor for the nucleon fields, $Q = \text{diag}(e, 0)$ is the nucleon charge matrix, $A^\mu$ and $F^{\mu\nu} = \partial_\mu A_\nu - \partial_\nu A_\mu$ are the usual photon field and field strength tensor, $m_N = 938.9187125(21)$ MeV is the average nucleon mass, $g_A = 1.2701(25)$ is the axial charge of the nucleon, $f_\pi = 92.21(14)$ MeV is the pion decay constant, $\pi^a$ is an isovector of pion fields, the $\tau^a$ are Pauli matrices acting in isospin space, and $\Delta m_N = m_n - m_p = 1.2933322(4)$ MeV is the nucleon mass splitting \cite{Beringer:1900zz}.  
The last term of \eq{LNpigamma} is allowed by symmetries and power counting and must therefore be included in the Lagrangian. The charge symmetry odd $\kappa^{\fs{s}}$ and $\kappa^{\fs{v}}$ parametrize the strength of the unconstrained CSB nucleon-photon contact interaction as
\begin{equation}
	\kappa_{CT}^{CSB} \equiv \kappa^{\fs{s}} - \kappa^{\fs{v}}.
\end{equation}
The only pieces of the pion-photon Lagrangian needed for our purposes are
\begin{eqnarray}
	\mathcal{L}_{\pi\gamma} &=& \frac{1}{2}\partial_\mu\pi^a\partial^\mu\pi^a - \frac{m_\pi^2}{2}\pi^a\pi^a- \frac{1}{4}F_{\mu\nu}F^{\mu\nu} + eA^\mu\partial_\mu\pi^a\pi^b\varepsilon^{3ab}.
	\label{Lpigamma}
\end{eqnarray}
where $m_\pi = 139.57018(35)$ MeV is the (charged) pion mass \cite{Beringer:1900zz}. 
We note that the isospin violating $\pi^\pm-\pi^0$ mass difference does not violate charge symmetry and can be ignored to leading order in chiral perturbation theory.

Using standard techniques\footnote{Some tedious Dirac algebra can be simplified by noting that expressions involving the pseudovector couplings of the chiral Lagrangian can be reduced to expressions involving only pseudoscalar couplings through the light front coordinate identity $\frac{1}{\fs{k}-m} = \sum_s \frac{u(k,s)\bar{u}(k,s)}{k^2-m^2} + \frac{\gamma^+}{2k^+}$. The second term does not contribute to the form factors in either diagram. This technique is used for instance in Ref.~\cite{Miller:2002ig}.} we can express the CSB form factors of  Eq.~\eqref{defFvs} as
\begin{subequations}\label{CSBterms}\begin{align}
	F_1^{CSB}(Q^2) =&  -2\left( \frac{g_A m_N}{f_\pi} \right)^2\left(\tilde{I}_1(Q^2,m_p,m_n) - \tilde{I}_1(Q^2,m_n,m_p)\right)\\\nonumber
	&\hspace{20pt}+ \left( \frac{g_A m_N}{f_\pi} \right)^2\left(\tilde{J}_1(Q^2,m_p,m_n)-\tilde{J}_1(Q^2,m_n,m_p)\right)\\
	F_2^{CSB}(Q^2) =&  4\left( \frac{g_A m_N}{f_\pi} \right)^2 \left(I_2(Q^2,m_p,m_n) - I_2(Q^2,m_n,m_p)\right)\\\nonumber
		&\hspace{20pt} + 2\left( \frac{g_A m_N}{f_\pi} \right)^2\left(J_2(Q^2,m_p,m_n)-J_2(Q^2,m_n,m_p)\right)\\\nonumber
		&\hspace{20pt} + \kappa_{CT}^{CSB},
\end{align}\end{subequations}
where the integrals $I_I$ arise from the photon hitting the intermediate nucleon as in Fig.~\ref{diagrams}(b), and the integrals $J_i$ arise from the photon hitting the pion as in Fig.~\ref{diagrams}(a). The results are:
\begin{subequations}\label{integrals}\begin{align}
	I_1(Q^2,m_E,m_I) &= \int_0^1 dx\int \frac{d^2k_\perp}{2(2\pi)^3} \frac{x\left[ \v{k}_\perp^2 - \frac{1}{4}x^2Q^2 + x^2m_E^2 + 2xm_E(m_I-m_E) \right]}{D_N^+(\v{k}_\perp) D_N^-(\v{k}_\perp)},\\\nonumber\\
	J_1(Q^2,m_E,m_I) &= \int_0^1 dx \int\frac{d^2k_\perp}{(2\pi)^3} \frac{x\left[ \v{k}_\perp^2 - \frac{1}{4}(1-x)^2Q^2 + x^2m_E^2 +2xm_E(m_I-m_E) \right]}{D_\pi^+(\v{k}_\perp) D_\pi^-(\v{k}_\perp)},\\\nonumber\\  
	I_2(Q^2,m_E,m_I) &=   \int_0^1 dx \int\frac{d^2k_\perp}{2(2\pi)^3} \frac{x^3m_E^2 + x^2 m_E(m_I-m_E)}{D_N^+(\v{k}_\perp) D_N^-(\v{k}_\perp)},\\\nonumber\\
	J_2(Q^2,m_E,m_I) &= \int_0^1 dx \int\frac{d^2k_\perp}{(2\pi)^3} \frac{x^2(1-x)m_E^2 + x(1-x)m_E(m_I-m_E)}{D_\pi^+(\v{k}_\perp)D_\pi^-(\v{k}_\perp)}.
\end{align}\end{subequations}
In these integrals $m_I$ and $m_E$ denote masses of internal and external nucleons respectively. We denote $Q^2=0$ subtractions with a tilde, i.e. $\tilde{I}_1(Q^2,m_E,m_I) \equiv I_1(Q^2,m_E,m_I) - I_1(0,m_E,m_I)$. The denominator factors above are given by
\begin{eqnarray}
	D_N^\pm(\v{k}_\perp) &=& (\v{k}_\perp \pm \frac{1}{2}x\v{q}_\perp)^2 + x^2m_E^2 +x(m_I^2-m_E^2) + (1-x)m_\pi^2\\\nonumber
	D_\pi^\pm(\v{k}_\perp) &=& (\v{k}_\perp \pm \frac{1}{2}(1-x)\v{q}_\perp)^2 + x^2m_E^2 +x(m_I^2-m_E^2) + (1-x)m_\pi^2.
	\label{denom}
\end{eqnarray}
The renormalization condition $F_1^{p}(0) = 1$ dictates $F_1^{CSB}(0) = 0$. We perform integrals of $\v{k}_\perp$ analytically with dimensional regularization and perform the remaining integrals over $x$ and a Feynman parameter numerically.

\subsubsection{Interpretation}
Some features of these results are noteworthy. The large CSB effect found by KL arises from $F_2^{CSB}$ and is driven by a logarithmically divergent  term that, in accordance with resonance saturation prescriptions, is cut off at the vector meson mass $m_V$ to give a contribution containing $\log m_V/m_\pi$. This means that the contact term used by KL includes a renormalization scale-dependent counterterm.

Our relativistic  expression for  $F_2^{CSB}$ is a convergent integral. One can see this by taking $Q^2=0$ and carrying out the integral over  $\v{k}_\perp$. One then obtains a term of the form $\log m_N/m_\pi$ with the same coefficient as that of KL. Since $m_N$ and $m_V$ are numerically similar the two approaches give similar results. However, our RB$\chi$PT calculation only includes renormalization-scale independent loop contributions to $F_2^{CSB}$ and therefore does not introduce renormalization scale-dependence into the contact term of Fig.~\ref{diagrams}(c).

\subsection{Including Form Factors}\label{vff}

The NLO contributions to $G_M^{CSB}$ computed in Refs.~\cite{Lewis:1998iu, Kubis:2006cy} give significant corrections to LO results.
Therefore it is useful to find a way to estimate the effects of  further higher-order corrections.
It is natural  to use 
  phenomenological vertex form factors  for this purpose because these  functions take  resummations of infinite classes of vertex corrections into account. Pion electroproduction measurements provide a $\pi\pi\gamma$ form factor~\cite{Blok:2008jy,Huber:2008id}
\begin{equation}
	F_\pi(Q^2) = \frac{1}{1 + Q^2/\Lambda_\pi^2},\hspace{20pt} \Lambda_\pi = 677 \pm 16 \text{ MeV},
	\label{Fpi}
\end{equation}
where $q^\mu$ is the momentum carried by the photon at the vertex and   $Q^2 = -q^2$. This form factor can be included by simply multiplying the integrals $J_1$ and $J_2$ by $F_\pi(Q^2)$ in Eq.~\eqref{CSBterms}.

More subtleties arise in constructing a $\pi NN$ form factor. Precise measurements of the pion-nucleon coupling constant are difficult, see for example Ref.~\cite{Stoks:1992ja}, and we will not attempt to extract a $\pi\pi N$ form factor directly from pion-nucleon scattering measurements. Instead we consider the nucleon axial current matrix element, which has  been accurately measured in neutrino scattering experiments. Chiral symmetry dictates that this matrix element is parametrized by axial vector and pseudoscalar form factors $G_A$ and $G_P$. Neutrino scattering measurements do not distinguish between axial vector and pseudoscalar effects, and so for analysis of these measurements the axial current matrix element as a whole is parametrized in terms of $G_A$ through a PCAC (tree level $\chi$PT) relation as\footnote{Since we are considering charge symmetry breaking in this paper we should in principle include a CSB tensor operator usually called a second class current. The point is that the experimental measurements discussed do not distinguish between axial, induced pseudoscalar, and second-class currents and so the precise division is unimportant for our purposes.}
\begin{eqnarray}
	\mbraket{N(p^\prime)}{A^\mu_a}{N(p)} &=&  \bar{u}(p^\prime)\left[ \gamma^\mu G_A(Q^2) + \frac{(p^\prime - p)^\mu}{2m_N}G_P(Q^2) \right]\gamma_5 \frac{\tau_a}{2} u(p)\\\nonumber
	&\simeq& \bar{u}(p^\prime)\left[ \gamma^\mu G_A(Q^2) + \frac{2m_N (p^\prime - p)^\mu}{m_\pi^2 + Q^2}G_A(Q^2) \right]\gamma_5 \frac{\tau_a}{2}u(p)
	\label{axialMEapprox}
\end{eqnarray}
where $A^\mu_a$ is an isovector axial current. The matrix element of $\partial_\mu A^\mu_a$ is connected to the matrix element of a pseudoscalar source such as a pion field through a chiral Ward identity~\cite{Scherer:2002tk}. Comparing this pseudoscalar parametrization with the explicit divergence of the form above, we have
\begin{eqnarray}
	\mbraket{N(p^\prime)}{\partial_\mu A^\mu_a}{N(p)} &=& \frac{ i m_\pi^2 f_\pi}{m_\pi^2 + Q^2}G_{\pi NN}(Q^2)\bar{u}(p^\prime)\gamma_5 \tau_a u(p)\\\nonumber
	&\simeq& \frac{im_N m_\pi^2}{m_\pi^2 + Q^2}G_A(Q^2)\bar{u}(p^\prime)\gamma_5\tau_a u(p).
\end{eqnarray}
Analysis of nucleon-neutrino scattering experiments shows that the axial current matrix element is well approximated by a dipole parametrization. With the above relation this parametrization gives a $\pi NN$ form factor~\cite{Budd:2003wb}
\begin{equation}
	G_{\pi NN}(Q^2) \simeq \frac{m_N}{f_\pi}G_A(Q^2) = \frac{g_A m_N}{f_\pi}\left( \frac{1}{1+Q^2/M_A^2} \right)^2,\hspace{20pt} M_A = 1.00 \pm 0.02 \text{ GeV}.
	\label{GpiNN}
\end{equation}
This parametrization reduces to the Goldberger-Treiman relation $G_{\pi NN}(0) = g_A m_N/f_\pi = 12.93$ at $Q^2=0$ and predicts $g_{\pi NN}\equiv G_{\pi NN}(-m_\pi^2) = 13.39 \pm 0.02 $. This is consistent with measurements of the Goldberger-Treiman discrepancy~\cite{Koch:1980ay}.

Immediately including $G_{\pi NN}$ as a vertex form factor would lead to the unacceptable prediction that electromagnetic neutrality of the neutron is violated by higher-order vertex corrections. Other higher-order effects must be included to restore gauge invariance. Fig.~\ref{diagrams}(a) only receives non-vanishing contributions when the internal nucleon propagator is on-shell,\footnote{This can be easily seen in light-front coordinates. Unless the momentum $k_\pi$ of the internal pion satisfies $0 < k_\pi^+ < p^+$ all of the $k_\pi^-$ poles are on the same half of the complex plane and the $k_\pi^-$ integral vanishes. For $0<k_\pi^+<p^+$ closing the $k_\pi^-$ integral in the upper half plane picks out the pole at $(p-k_\pi)^2=m_I^2$. In Fig.~\ref{diagrams}(b) analogous reasoning shows the internal pion propagator is placed on shell with $(k_N-p)^2=m_\pi^2$.} and so we take $G_{\pi NN}$ as our vertex form factor for this diagram. Demanding that the sum of Fig.~\ref{diagrams}(a) and Fig.~\ref{diagrams}(b) preserves gauge invariance constrains the $\pi NN$ form factor included in Fig.~\ref{diagrams}(b) to be identical to the $\pi NN$ form factor in (a) when both are expressed as functions of their corresponding integration variables $x$, $\v{k}_\perp$.

The loop momentum variables of the two diagrams are simply related in the Drell-Yan-West frame, in which
\begin{equation}
  q = (q^+, q^-,\v{q}_\perp) = \left( 0, \frac{Q^2}{p^+}, \v{q}_\perp \right),\hspace{20pt} p=(p^+,p^-,\v{p}_\perp) = \left( p^+, \frac{m_E^2}{p^+},\v{0}_\perp \right)
\end{equation}
where $q^+ = q^0 + q^3$, $q^- = q^0 - q^3$, etc. express momenta in light-front coordinates. Kinematic simplifications in this frame allow us to express form factors for both diagrams in terms of the corresponding loop momentum momentum $k_\pi$ or $k_N$ or in terms the common integration variables $x, \v{k}_\perp$. Noting that in Eq.~\eqref{integrals} the integration variable $x$ is defined as $k_\pi^+ = xp^+$ for Fig.~\ref{diagrams}(a) and $k_N^+ = (1-x)p^+$ for Fig.~\ref{diagrams}(b), our $\pi NN$ vertex form factor is given by
\begin{eqnarray}\label{FpiNN}
	F_{\pi NN}(x,\v{k}_\perp) &=& \left( \frac{M_A^2}{M_A^2 + \frac{\v{k}_\perp^2}{1-x} - xm_E^2 + \frac{x}{1-x}m_I^2} \right)^2 \label{pinff}\\\nonumber\\\nonumber
	&=& \begin{dcases} \left( \frac{M_A^2}{k_\pi^2 - M_A^2} \right)^2 & \text{Fig.~\ref{diagrams}(a)} \\
		\left( \frac{\left(\frac{1-x}{x}\right)M_A^2}{k_N^2 - m_I^2 - \frac{1-x}{x}(M_A^2 - m_\pi^2)} \right)^2 & \text{Fig.~\ref{diagrams}(b)}. \end{dcases}
\end{eqnarray}
Form factors for each vertex $F_{\pi NN}(x,\v{k}_\perp)F_{\pi NN}(x,\v{k}_\perp + \v{q}_\perp)$ should be inserted in the three dimensional integrals of Eq.~\eqref{integrals} since inserting them in the original four dimensional loop integrals adds unphysical propagator poles. 
 Our procedure will be to use the difference between results obtained using the form factor of \eq{pinff} and unity as a measure of the uncertainty in higher order corrections. This means that we define the error associated with the form factor to be plus or minus the difference between using and not using the form factor.

\subsection{Resonance Saturation}\label{vm}

A predictive calculation of the CSB form factors requires a model estimate of the unconstrained counterterm $\kappa_{CT}^{CSB}$. The resonance saturation technique provides such a model estimate~\cite{Ecker:1988te}. Resonance saturation involves adding heavier resonance fields to an EFT and identifying contributions from unconstrained operators in the EFT with contributions from resonance operators in the extended theory. 

Resonance saturation assumes that unknown LECs encoding physics beyond the original EFT are well-approximated by the coefficients resulting from integrating out a set of resonance fields. There is no consistent power counting scheme for loop-level resonance effects, and only tree-level resonance exchange is typically included in resonance saturation estimates. The dominance of tree-level contributions is supported by for example large $N_c$ arguments, but loop-level contributions are not parametrically suppressed within the chiral expansion~\cite{Bernard:2007zu}. There may be ultraviolet physics encoded in LECs that is not captured by a model of tree-level resonance exchange, and resonance saturation estimates are not guaranteed to accurately reproduce LECs up to parametrically small errors.

The resonance saturation technique has been shown to work well in practice for many LECs in mesonic $\chi$PT and in HB$\chi$PT. Nucleon magnetic moments were shown in~\cite{Bernard:1996gq} to be well saturated by the effects of $\rho$ and $\omega$ resonance exchange in accordance with the idea of vector meson dominance. Nucleon isoscalar electric and magnetic radii are also well-described by vector meson resonance effects, though isovector radii are not accurately described~\cite{Kubis:2000zd}. It is possible that CSB contributions from $\rho$ and $\omega$ mesons will saturate electromagnetic radii better than isospin conserving radii because contributions from heavier resonances are suppressed by additional powers of resonance masses in the CSB case~\cite{Kubis:2006cy}.

\begin{figure}
	\includegraphics[width=.7\textwidth, height=.15\textheight]{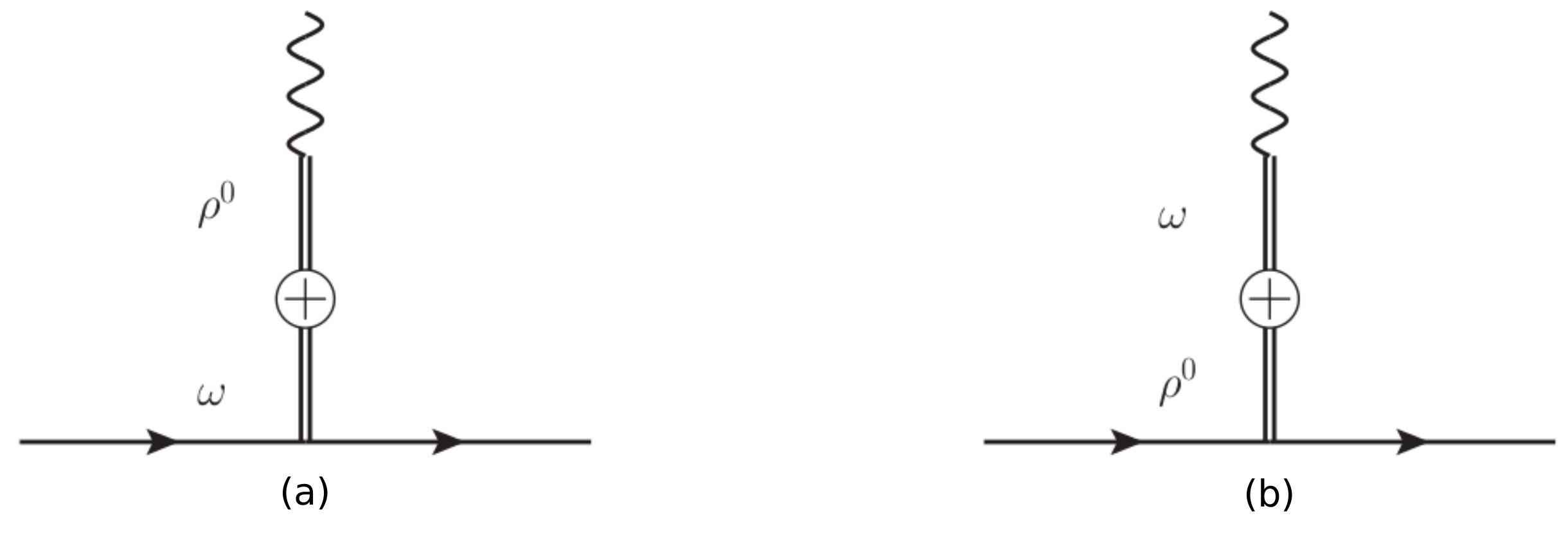}
	\caption{CSB contributions to the proton's neutral weak form factors from tree-level resonance exchange. CSB effects arise from mixing between the isoscalar $\omega$ and isovector $\rho$. The mixing vertex $\Theta_{\rho\omega}$ is denoted by a crossed circle. These diagrams provide a resonance saturation estimate for the unconstrained contact interaction shown in Fig.~\ref{diagrams}(c).}
	\label{vmdiagrams}
\end{figure}

The lightest resonance contributions to $G^{CSB}$ arise from mixing between the isovector $\rho$ and isoscalar $\omega$ mesons. Tree-level diagrams describing this process are shown in Fig.~\ref{vmdiagrams}. The vector mesons carry the small momentum $q^\mu$ and so the meson-nucleon couplings can be organized in a derivative expansion. It is convenient to represent interacting, massive spin 1 fields with antisymmetric tensors~\cite{Ecker:1988te}. With this representation, the leading contributions to $F_2^{CSB}$ and $F_1^{CSB}$ arise from meson-nucleon couplings with zero and one derivative respectively. The Lagrangians describing these interactions and the leading meson-photon couplings are
\begin{eqnarray}
	\mathcal{L}_{NV} &=& -\frac{1}{2}\bar{N}\left[ \left( \frac{g_\rho}{m_V}\gamma^\mu\partial^\nu + \frac{g_\rho \kappa_\rho m_V\sigma^{\mu\nu}}{4m_N} \right) \rho_{\mu\nu}^a\tau^a \right] N\\\nonumber
	&&\hspace{20pt} - \frac{1}{2}\bar{N}\left[ \left( \frac{g_\omega}{m_V}\gamma^\mu\partial^\nu + \frac{g_\omega\kappa_\omega m_V\sigma^{\mu\nu}}{4m_N}\right)\omega_{\mu\nu} \right]N\\\nonumber
	\mathcal{L}_{V\gamma} &=&  - e \left(F_\rho\rho^0_{\mu\nu} + F_\omega \omega_{\mu\nu}\right) \partial^\nu A^\mu,
	\label{vmL}
\end{eqnarray}
where $F_\omega = 45.7$ MeV and $F_\rho = 152.5$ MeV are vector meson decay constants, $m_V \sim m_\rho = 770$ MeV is the chiral limit vector meson mass, and $g_\rho $, $g_\omega$, $\kappa_\rho$, and $\kappa_\omega$ are coupling constants discussed in the next section. With these interactions and the empirical $\rho-\omega$ mixing amplitude given by~\cite{Kucukarslan:2006wk}
\begin{equation}
	\Theta_{\rho\omega} = (-3.75\pm0.36)\times 10^{-3}\text{ GeV}^2,
\end{equation}
the CSB form factor contributions from $\rho-\omega$ mixing are given by
\begin{eqnarray}
	F_1^{CSB}(Q^2)_{\rho-\omega} &=& (g_\rho F_\omega - g_\omega F_\rho )\frac{\Theta_{\rho\omega}Q^2}{m_V(m_V^2 + Q^2)^2},\\\nonumber
	F_2^{CSB}(Q^2)_{\rho-\omega} &=& (g_\omega\kappa_\omega F_\rho - g_\rho\kappa_\rho F_\omega)\frac{m_V\Theta_{\rho\omega}}{(m_V^2 + Q^2)^2},
	\label{kappamix}
\end{eqnarray}
in agreement with KL. Details of computing Feynman diagrams with antisymmetric tensor fields are given in~\cite{Ecker:1988te}. 
The resonance scale $m_V$ was  used by KL as the renormalization scale when matching the renormalization scale-dependent contact term in HB$\chi$PT with the scale independent resonance estimate. Our RB$\chi$PT approach gives a renormalization scale-independent result for $F_2^{CSB}$ and there is no choice of what scale should be used in matching to the resonance saturation estimate.

\subsubsection{Strong coupling constants}\label{couplings}
As noted above,
Ref.~~\cite{Wang:1900ta}  takes  takes  issue with the work of KL~\cite{Kubis:2006cy} for using very large values of vector-meson nucleon coupling constants. Their values are given in Table~\ref{parameters}, along with two other sets~\cite{Dumbrajs:1983jd,Ericson:1988gk} and ~\cite{Coon:1987kt}.  KL obtained (using NLO HB$\chi$PT)  $G_M^{CSB}(0) = 0.025 \pm 0.02$  in which the large error bar  is almost entirely due to uncertainty in dispersive extractions of $\kappa_\omega = 0.37 \pm 0.21$. 
The effect of this uncertainty in $\kappa_\omega$ is enhanced by the large value $g_\omega \sim 42$ that KL take from the dispersion analysis of Refs.~\cite{Belushkin:2006qa, Hammer:2003ai, Mergell:1995bf}. 

However, there has been a long standing controversy between values of $g_\omega$ determined from electromagnetic form factors and those obtained from nucleon-nucleon scattering and quark models. Large variations between different determinations of $g_\rho$ and $g_\omega$ in particular are clearly visible in Table~\ref{parameters}. We take the textbook values of~\cite{Ericson:1988gk} to be definitive.  These are taken from a compilation~\cite{Dumbrajs:1983jd} which is the summary 
of several years of work.  The value of $g_\omega$ is close to the value obtained using SU(3) symmetry~\cite{Nagels:1977ze}. This is also close to  the   ones used in 
calculations of nuclear CSB effects~\cite{Gardestig:2004hs, Nogga:2006cp}, and are typical of those used in nucleon-nucleon potentials computed in the  one-boson exchange approximation which is relevant here. The value of $g_\omega$ is  10.1 instead of the large value of 42 used by KL. 
      
There is another way to determine the strong coupling constants appropriate for calculations of CSB effects. This is to take them from the most accurate information regarding CSB in nucleon-nucleon scattering (the difference between electromagnetically corrected $pp$ and $nn$ $^1S_0$ scattering lengths) and the binding energy difference between mirror nuclei (in particular the pair $^3$He-$^3$H). The  principle cause of the $^3$He-$^3$H binding energy difference, $\sim760$ KeV, is the Coulomb interaction and other purely electromagnetic effects. It is well established that  $ 73\pm 22$ KeV of the  difference of the $^3$He-$^3$H  binding energies  arises from CSB effects in the strong interaction, and that the $nn$ $^1S_0$ 
scattering length is more attractive than the one for $pp$ by 1.5 $\pm0.5 $ fm , see  {\textit e.g.} the reviews of Refs.~\cite{Miller:1990iz,Miller:1994zh,Miller:2006tv}.

Coon \& Barrett~\cite{Coon:1987kt} showed that the potential $V_{\rho-\omega}$ obtained from the exchange of an isospin-mixed $\rho-\omega$ system between nucleons could account for both phenomena. This finding was confirmed by several authors~\cite{Miller:1990iz,Miller:1994zh,Miller:2006tv}. However, there are other effects that lead to scattering length and binding energy
differences  such as baryon mass differences in two-boson exchange potentials and the effects of $\pi-\eta$ mixing. All of these effects are controlled by the small mass difference between down and up quarks $m_d-m_u>0$ and all have the same sign. Thus we take Coon \& Barrett's resonance parameters, which saturate experimental bounds on strong CSB effects with $\rho-\omega$ mixing contributions alone, to represent a \emph{maximum} upper limit on $\rho-\omega$ mixing contributions to CSB contact interactions. The measured value of $\Theta_{\rho\omega}$ has decreased by about 20\% since the  1987 work of Coon \& Barrett. With the modern value of $\Theta_{\rho\omega}$, the textbook value of $\kappa_\omega = 0.1\pm 0.2$, and the other coupling constants denoted as Ref.~\cite{Coon:1987kt} in Table~\ref{parameters}, the upper limit provided by $^3$He-$^3$H binding energy difference becomes $g_\omega = 19\pm 5$. The uncertainty shown on this estimate corresponds to the change in maximum $g_\omega$ allowed when $\kappa_\omega$ and $\Theta_{\rho\omega}$ are varied across their confidence intervals shown. 
  
To summarize we default to the coupling constants of Refs.~\cite{Dumbrajs:1983jd,Ericson:1988gk}. We also consider the couplings of Ref.~\cite{Coon:1987kt}, adjusted for modern $\Theta_{\rho\omega}$ measurements and to include the spread $\kappa_\omega = 0.1\pm 0.2$ rather than fixing $\kappa_\omega$ at 0, as an upper limit on the size of $\rho-\omega$ mixing contributions. For comparison we show some results for KL's choices in Ref.~\cite{Kubis:2006cy}, also shown in Table~\ref{parameters}. We will denote which of these three sets of coupling constants are used to calculate various results below by their corresponding value of $g_\omega = 10$~\cite{Dumbrajs:1983jd,Ericson:1988gk}, 19~\cite{Coon:1987kt}, 42~\cite{Kubis:2006cy}.

\section{Results}\label{results}

\setlength{\tabcolsep}{10pt}

\begin{table}
	\centering
	\begin{tabular}{|c|c||c|c|}\hline
	$m_N$ & 938.92 MeV & $\Delta m_N$ & 1.29 MeV\\\hline
	$m_\pi$ & 139.57 MeV & $g_A m_N/f_\pi$ & 12.93\\\hline
	$\Lambda_\pi$ & 677 MeV & $M_A$ & 1.00 GeV \\\hline
	$F_\rho$ & 152.5 MeV & $F_\omega$ & 45.7 MeV\\\hline
	$m_V$ & 770 MeV & $\Theta_{\rho\omega}$ & $(-3.75 \pm 0.36)\times 10^{-3}$ GeV$^2$ \\\hline
	$g_\rho$~\cite{Kubis:2006cy}  & 5.2 & $g_\omega$~\cite{Kubis:2006cy} & 42\\\hline
	$\kappa_\rho$~\cite{Kubis:2006cy} & 6.0 & $\kappa_\omega$~\cite{Kubis:2006cy} & $ 0.21 \pm 0.37$ \\\hline
$g_\rho$~\cite{Dumbrajs:1983jd,Ericson:1988gk}  & $2.6\pm0.14$& $g_\omega$~\cite{Dumbrajs:1983jd,Ericson:1988gk} & $10.1\pm 0.8$\\\hline
	$\kappa_\rho$~\cite{Dumbrajs:1983jd,Ericson:1988gk} & $6.1 \pm0.4$ & $\kappa_\omega$~\cite{Dumbrajs:1983jd,Ericson:1988gk}  & $ 0. 1 \pm 0.2$ \\\hline	
	$g_\rho$~\cite{Coon:1987kt} & 5.5& $g_\omega$~\cite{Coon:1987kt} & $16.2\pm 0.46\,\quad (19\pm 5) $\\\hline
	$\kappa_\rho$~\cite{Coon:1987kt} & $6.6 \pm0.4$ & $\kappa_\omega$~\cite{Coon:1987kt}  &\hspace{20pt} $0\,\hspace{20pt} (0.1\pm0.2)$ \\\hline
	\end{tabular}
	\caption{Parameter values used for numerical evaluation. The details are explained in the   text.}
	\label{parameters}
\end{table}
\begin{figure}
	\begin{center}
	\includegraphics[scale=0.415]{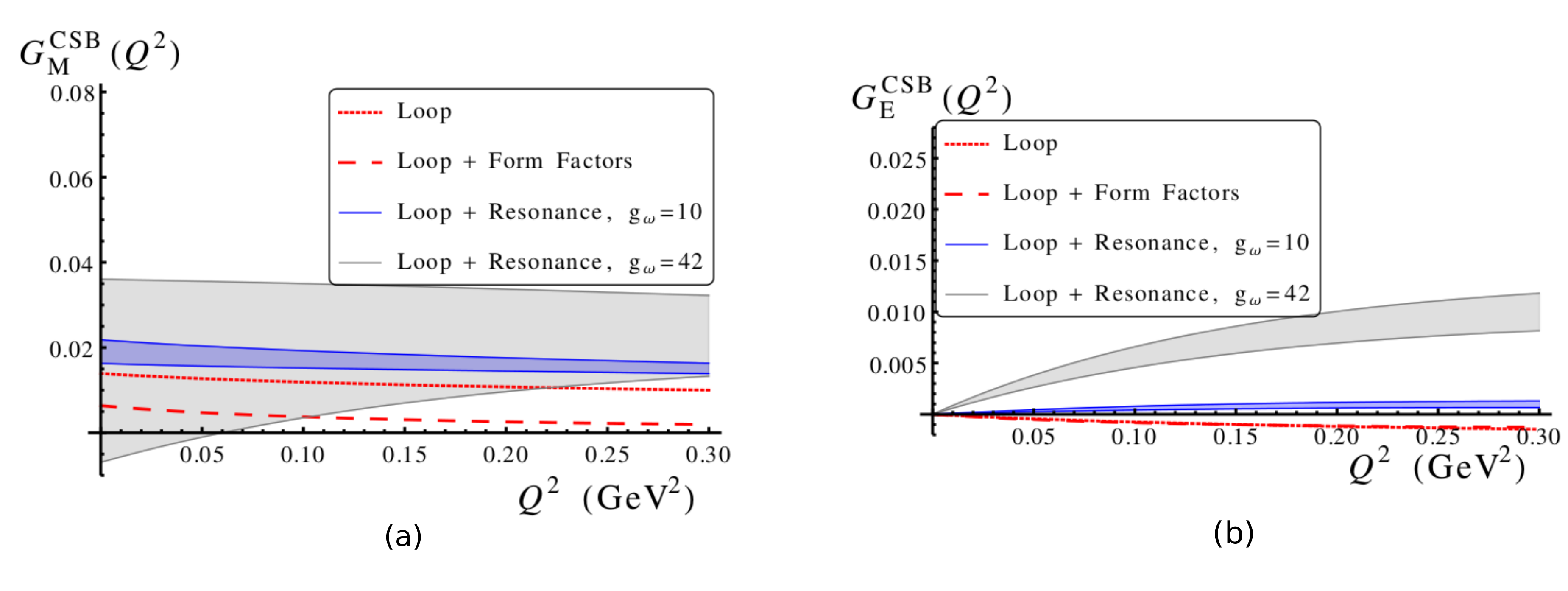}
	\end{center}
	\caption{(Color Online) The CSB form factors $G_M^{CSB}(Q^2)$ (a) and $G_E^{CSB}(Q^2)$ (b) contributing to the proton's neutral weak form factors. The darker blue shaded regions show our leading-order prediction for the form factors. The unconstrained contact term $\kappa_{CT}^{CSB}$ is estimated with resonance saturation and the coupling choice $g_\omega = 10$ incorporates the nuclear scattering constraints described in the text  and Table~\ref{parameters} from~\cite{Dumbrajs:1983jd,Ericson:1988gk}. The width of the shaded regions results from uncertainty in the resonance parameters $\kappa_\omega$ and $\Theta_{\rho\omega}$. 
The lighter gray shaded regions show the same results with the coupling $g_\omega = 42$ taken by KL from dispersion analysis. The dotted red lines show the loop contributions only (no contact terms) calculated in RB$\chi$PT, and the dashed red lines show the loop contributions when phenomenological $\pi \pi \gamma$ and $\pi NN$ vertex form factors are included.} 
	\label{csbvmplot}
\end{figure}

\begin{figure}[h]
	\begin{center}\
	  \includegraphics[scale=0.44]{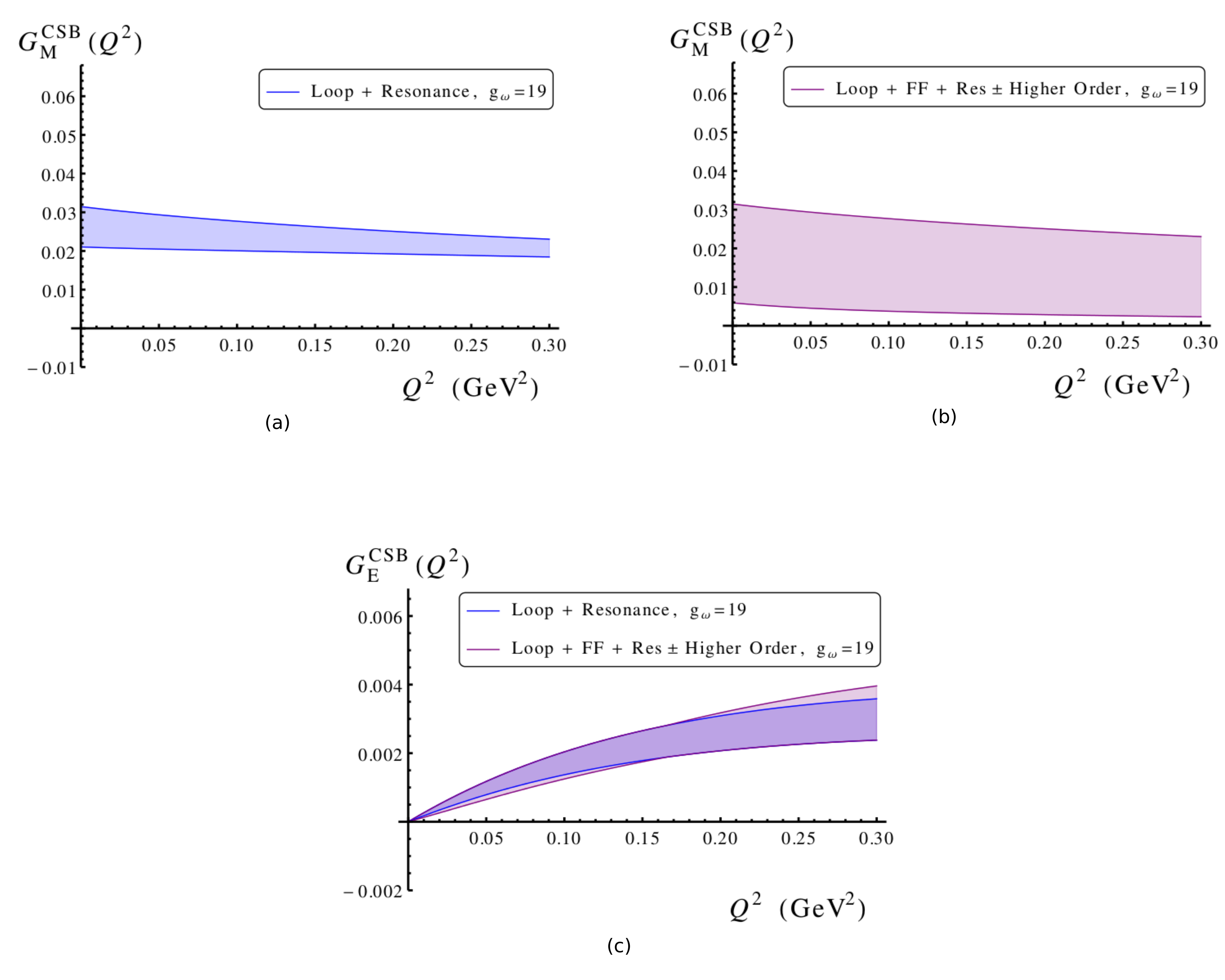}
	\end{center}
	\caption{(Color Online) The CSB form factor contributions including uncertainty estimates for resonance parameters and for contributions beyond LO. The resonance parameters chosen correspond to upper limit on $\rho-\omega$ mixing contributions consistent with nuclear scattering and binding energy measurements as described in the main text. The darker blue regions in (a) and (c)  include only resonance uncertainty.  The lighter red regions in (b) and (c) additionally include an estimate of higher-order term uncertainty as characterized by phenomenological vertex form factors. This estimate is found by considering $G^{CSB} \pm |\Delta G^{CSB}|$, where $\Delta G^{CSB}$ is the difference between the CSB form factors with and without phenomenological vertex form factors. This estimate of higher-order uncertainty adds significant uncertainty to $G_M^{CSB}$ but negligible uncertainty to $G_E^{CSB}$.}
	\label{csbvmvffplot}
\end{figure}

\begin{figure}[h]
	\begin{center}\
	  \includegraphics[scale=0.44]{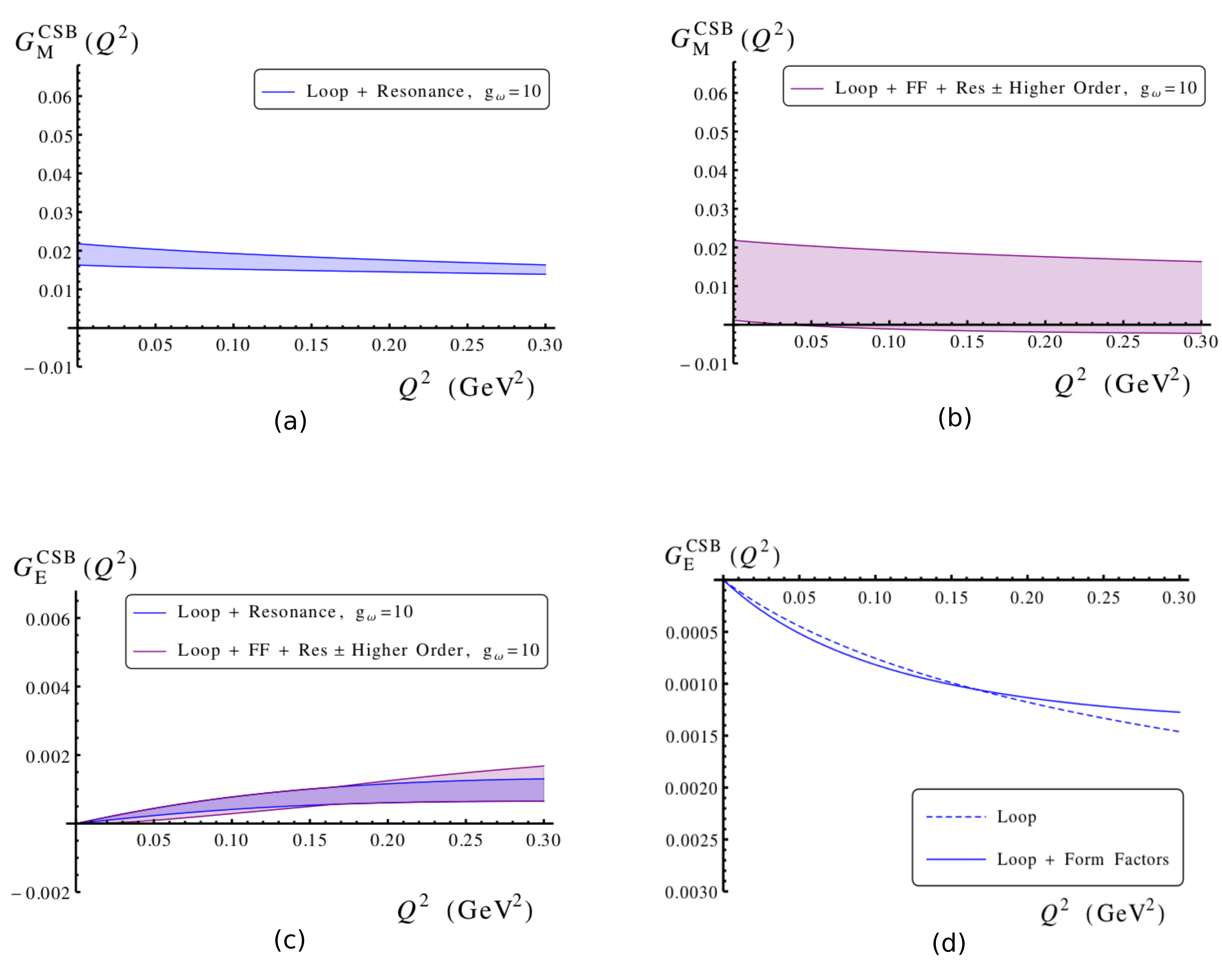}
	\end{center}
	\caption{(Color Online) The CSB form factor contributions (using coupling constants of~\cite{Dumbrajs:1983jd,Ericson:1988gk}) including uncertainty estimates for resonance parameters and for contributions beyond LO. The darker blue regions in (a) and (c)  include only resonance uncertainty and are identical to the blue regions in Fig.~\ref{csbvmplot}.  The lighter red regions in (b) and (c) additionally include an estimate of higher-order term uncertainty as characterized by phenomenological vertex form factors as above. Finally (d) shows only the loop contributions to $G_E^{CSB}$. These provide a leading-order $\chi$PT prediction for $G_E^{CSB}$ that is numerically dominated by the formally next-to-leading order resonance contributions included in (c).}
	\label{final}
\end{figure}

Our first set of results is shown in Fig.~\ref{csbvmplot}.   Our LO results for the CSB form factors include both the loop contributions of Eq.~\eqref{CSBterms} and the counterterm contribution estimated by resonance saturation. 
Therefore 
the full uncertainty of our calculation includes both uncertainty in resonance parameters and uncertainty about neglected higher-order effects. The form factors in Eq.~\eqref{Fpi} and Eq.~\eqref{FpiNN} describe phenomenologically relevant effects of all orders in the chiral expansion. The change in the CSB magnetic moment when phenomenological vertex form factors are included, $\Delta G^{CSB}_M(0) = -0.008$, is of comparable magnitude but opposite sign to the NLO corrections to $G_M^{CSB}(0)$ calculated in HB$\chi$PT~\cite{Lewis:1998iu,Kubis:2006cy}. We take $|\Delta G^{CSB}(Q^2)|$, the magnitude of the difference between the CSB form factors with and without phenomenological vertex form factors, to roughly characterize the size of higher-order corrections to the LO result. 

The first notable point from the results shown in Fig.~\ref{csbvmplot} is that our RB$\chi$PT results for the pion loop term (red line) agree with HB$\chi$PT results. Thus we substantiate the finding that the pion loop contributions are about 10 times larger than expected from na{\"i}ve dimensional analysis. The major differences between our results and those of KL are due to the differences in the strong coupling constants used. With more modest vector meson-nucleon coupling constants, the uncertainty in resonance contributions to $G_M^{CSB}$ found by KL is greatly reduced. The overall size and uncertainty of $G_E^{CSB}$ are both reduced by using smaller coupling constants. The constraint $G_E^{CSB}(Q^2=0)=0$ ensures that this function mainly depends on the small loop momentum region and the effects of introducing form factors are negligible. Introducing form factors does reduce $G_M^{CSB}(Q^2)$. 

The next set of results are shown in Fig.~\ref{csbvmvffplot} and explore the effects of using the upper limit coupling constants taken from Coon \& Barrett~\cite{Coon:1987kt}. The adjusted upper limit $g_\omega =19$ is between the values $g_\omega =10$ and $g_\omega=42$ shown above. This figure demonstrates the strong dependence of our results on the value chosen for $g_\omega$ and should allow the reader to interpolate in the event that more definitive coupling constant determinations become available (note that $g_\rho$ is comparable in these results and the $g_\omega = 42$ results in Fig.~\ref{csbvmplot} but somewhat smaller in the $g_\omega = 10$ results). Once again including form factors has little effect on $G_E^{CSB}$ but reduces the central value of $G_M^{CSB}$. Including form factors also allows us to quantify the uncertainty due to higher-order corrections to our LO results, as discussed above and in Sec.~\ref{vff}. Adding this additional measure of uncertainty leads to the broader error bands for results including form factors in Fig.~\ref{csbvmvffplot} and Fig.~\ref{final}.

The third set of results may be thought of as our final results, obtained  using the coupling  constants  of ~\cite{Dumbrajs:1983jd,Ericson:1988gk}. These are shown in Fig.~\ref{final}. Predictions for the CSB magnetic moment and charge and magnetic radii both with and without phenomenological form factors are shown in Table~\ref{moments}. Results are shown for both $g_\omega = 10$ and $g_\omega=19$.

It should be noted that the resonance saturation contact term estimate is necessary for a prediction of $G_M^{CSB}$ at LO\footnote{The HB$\chi$PT loop contribution is divergent and manifestly unphysical. The RB$\chi$PT contribution is finite but in either case physical predictions require both loop and counterterm contributions.} but does not contribute to $G_E^{CSB}$ until NLO. Regardless, the formally subleading resonance contributions to $G_E^{CSB}$ are numerically dominant. We therefore include the full $Q^2$ dependence of the resonance contributions in Figs.~\ref{csbvmvffplot} and \ref{final} and the resonance results of Table~\ref{moments}. For comparison we show the loop contributions to $G_E^{CSB}$ separately in Fig.~\ref{final}(d).


\begin{table}
	\centering
	\begin{tabular}{|l|| l| l | l|}\hline
	 & $G_M^{CSB}(0)$ & $\rho_M^{CSB}$ (fm$^2$) & $\rho_E^{CSB}$ (fm$^2$)\\\hline
	Loop& .014 & .0012 & .0004 \\\hline
	Loop + Form Factors & .006 & .0017 & .0009 \\\hline
	Loop + Resonance, $g_\omega$ =10 & .019 $\pm$ .003 &.0010 $\pm$ .0004 & -.0003 $\pm$ .0001 \\\hline
	Loop + FF + Res, $g_\omega=10$ &.012 $\pm$ .003 & .0006 $\pm$ .0004 & -.0007 $\pm$ .0001 \\\hline
	Loop + Resonance, $g_\omega$ = 19 & .026 $\pm$ .005 & .0012 $\pm$ .0007 & -.0009 $\pm$ .0002 \\\hline
	Loop + FF + Res, $g_\omega= 19$ & .019 $\pm$ .005  & .0009 $\pm$ .0007 & -.0013 $\pm$ .0002 \\\hline
\end{tabular}
\caption{Results for the CSB magnetic moment $G_M^{CSB}(0)$ and electric and magnetic radii $\rho^{CSB} = -\frac{dG^{CSB}}{dQ^2}(0)$ from RB$\chi$PT with resonance saturation. The first two lines show the loop contributions only, without and with phenomenological $\pi\pi\gamma$ and $\pi NN$ vertex form factors. The loop contribution to $\rho_E^{CSB}$ is a prediction of $\chi$PT at leading-order. The leading-order prediction for $G_M^{CSB}$ requires a resonance estimate of contact terms and is given by the Loop + Resonance line above. Formally next-to-leading resonance contributions to $\rho_E^{CSB}$ are numerically larger than the leading-order result and are included in the values for $\rho_E^{CSB}$ shown in the last four lines. The third and fourth lines take the resonance parameter choices of Refs.~\cite{Dumbrajs:1983jd,Ericson:1988gk}, while the fifth and sixth lines use an upper bound on the size of resonance parameters taken from Ref.~\cite{Coon:1987kt}, for discussion of these coupling constant choices see Sec. \ref{couplings}. See the main text for details and in particular see Sec.~\ref{sum} for comparison to HB$\chi$PT results.}
	\label{moments}
\end{table}

\section{Summary and Discussion}\label{sum}

Our principal result is that charge symmetry breaking effects are too small to influence the extraction of nucleon strangeness measurements from parity-violating electron-proton scattering experiments. Including both uncertainty in resonance parameters and higher-order term uncertainty quantified by the magnitude of form factor contributions, our LO predictions are $G_M^{CSB}(0) = 0.012 \pm 0.003 \pm 0.008$ (the second error represents higher-order term uncertainty and is equal to the difference between the first and second lines of Table II) and $|G_E^{CSB}| < 0.002$ for $Q^2 < 0.3$ GeV$^2$. Comparing these results with current experimental bounds on strangeness form factors $G_M^s = 0.33 \pm 0.4$, $G_E^s = 0.006 \pm 0.02$ at $Q^2 = 0.1$ GeV$^2$~\cite{Armstrong:2012bi}, we see that our CSB predictions are an order of magnitude smaller than current experimental error bars. 

The predictions of HB$\chi$PT with resonance saturation made by KL are  $G_M^{CSB}(0) = 0.025 \pm 0.02$ and $|G_E^{CSB}| < 0.01$ for $Q^2 < 0.03$ GeV$^2$ including resonance parameter uncertainty~\cite{Kubis:2006cy}. The much larger resonance parameter uncertainty in these results arises from using a large $\omega$-nucleon coupling constant $g_\omega \sim 42$ taken from dispersion analysis. Experimental measurements of the $^3$He-$^3$H binding energy difference constrain $g_\omega \lesssim 19 \pm 5$ when $\rho-\omega$ mixing is treated as a resonance contribution to HB$\chi$PT contact operators. This is still larger than determinations from nucleon scattering of $g_\omega = 10.1$~\cite{Dumbrajs:1983jd,Ericson:1988gk}. Taking $g_\omega = 10.1$, the prediction of HB$\chi$PT with resonance saturation becomes $G_M^{CSB}(0) = 0.030 \pm 0.003$ at NLO and $|G_E^{CSB}|<0.002$ at LO for $Q^2 < 0.03$ GeV$^2$. This is once again an order of magnitude smaller than current experimental uncertainties on nucleon strangeness.

Our results demonstrate good agreement between LO loop contributions in RB$\chi$PT and HB$\chi$PT. The RB$\chi$PT loop contribution of 0.014 to $G_M^{CSB}(0)$ agrees with the LO HB$\chi$PT loop contribution to better than 95\%. The RB$\chi$PT loop contribution to $\rho_M^{CSB}$ is smaller than the LO HB$\chi$PT loop contribution but larger than the loop contribution at NLO. The two frameworks therefore manifestly agree on $\rho_M^{CSB}$ up to higher-order corrections. The RB$\chi$PT loop contribution to $\rho_E^{CSB}$ is also smaller than the LO HB$\chi$PT loop contribution, but $\rho_E^{CSB}$ is numerically dominated by the resonance contribution in both frameworks and so we expect that differences can again be considered higher-order. 

RB$\chi$PT and HB$\chi$PT must give predictions for physical observables that agree up to higher-order errors once loop and counterterm contributions are included. It is encouraging to see that this agreement is achieved when using resonance saturation estimates for the counterterm contributions. A model independent chiral prediction for the CSB form factors still requires direct constraints on the CSB nucleon-photon contact interaction from experiment or QCD, but our investigations have found no reason to doubt the consistency of CSB form factor predictions using chiral loops and resonance saturation contact terms.

  \section*{Acknowledgements} This   work has been partially supported by 
U.S. D. O. E.  Grant No. DE-FG02-97ER-41014.  We thank U. van Kolck and M. Savage for useful discussions.

\bibliographystyle{h-physrev}
\bibliography{ChPT}

\begin{thebibliography}{10}

\bibitem{Armstrong:2012bi}
D.~Armstrong and R.~McKeown,
\newblock Ann.Rev.Nucl.Part.Sci. {\bf 62}, 337 (2012), 1207.5238.

\bibitem{Henley:1979ig}
E.~M. Henley and G.~A. Miller,
\newblock Mesons In Nuclei, Vol.I Edited by Rho M, Wilkinson D. Amsterdam, pp.
  405-434  (1979).

\bibitem{Miller:1990iz}
G.~A. Miller, B.~M.~K. Nefkens, and I.~Slaus,
\newblock Phys.Rept. {\bf 194}, 1 (1990).

\bibitem{Miller:1994zh}
G.~A. Miller and W.~T. Van~Oers,
\newblock Symmetries and fundamental interactions in nuclei Edited by Haxton,
  W.C., Henley, E.M., pp. 127-167  (1994), nucl-th/9409013.

\bibitem{Miller:2006tv}
G.~A. Miller, A.~K. Opper, and E.~J. Stephenson,
\newblock Ann.Rev.Nucl.Part.Sci. {\bf 56}, 253 (2006), nucl-ex/0602021.

\bibitem{Dmitrasinovic:1995jt}
V.~Dmitrasinovic and S.~Pollock,
\newblock Phys.Rev. {\bf C52}, 1061 (1995), hep-ph/9504414.

\bibitem{Miller:1997ya}
G.~A. Miller,
\newblock Phys.Rev. {\bf C57}, 1492 (1998), nucl-th/9711036.

\bibitem{Lewis:1998iu}
R.~Lewis and N.~Mobed,
\newblock Phys.Rev. {\bf D59}, 073002 (1999), hep-ph/9810254.

\bibitem{Kubis:2006cy}
B.~Kubis and R.~Lewis,
\newblock Phys.Rev. {\bf C74}, 015204 (2006), nucl-th/0605006.

\bibitem{Ecker:1988te}
G.~Ecker, J.~Gasser, A.~Pich, and E.~de~Rafael,
\newblock Nucl.Phys. {\bf B321}, 311 (1989).

\bibitem{Acha:2006my}
HAPPEX collaboration, A.~Acha {\em et~al.},
\newblock Phys.Rev.Lett. {\bf 98}, 032301 (2007), nucl-ex/0609002.

\bibitem{Paschke:2011zz}
K.~Paschke, A.~Thomas, R.~Michaels, and D.~Armstrong,
\newblock J.Phys.Conf.Ser. {\bf 299}, 012003 (2011).

\bibitem{Wang:1900ta}
P.~Wang, D.~Leinweber, A.~Thomas, and R.~Young,
\newblock Phys.Rev. {\bf C79}, 065202 (2009), 0807.0944.

\bibitem{GonzalezJimenez:2011fq}
R.~Gonzalez-Jimenez, J.~Caballero, and T.~Donnelly,
\newblock Phys.Rept. {\bf 524}, 1 (2013), 1111.6918.

\bibitem{Dumbrajs:1983jd}
O.~Dumbrajs {\em et~al.},
\newblock Nucl.Phys. {\bf B216}, 277 (1983).

\bibitem{Ericson:1988gk}
T.~E.~O. Ericson and W.~Weise,
\newblock (1988).

\bibitem{Weinberg:1978kz}
S.~Weinberg,
\newblock Physica {\bf A96}, 327 (1979).

\bibitem{Gasser:1983yg}
J.~Gasser and H.~Leutwyler,
\newblock Annals Phys. {\bf 158}, 142 (1984).

\bibitem{Gasser:1987rb}
J.~Gasser, M.~Sainio, and A.~Svarc,
\newblock Nucl.Phys. {\bf B307}, 779 (1988).

\bibitem{Jenkins:1990jv}
E.~E. Jenkins and A.~V. Manohar,
\newblock Phys.Lett. {\bf B255}, 558 (1991).

\bibitem{Kaplan:2005es}
D.~B. Kaplan,
\newblock (2005), nucl-th/0510023.

\bibitem{Scherer:2002tk}
S.~Scherer,
\newblock Adv.Nucl.Phys. {\bf 27}, 277 (2003), hep-ph/0210398.

\bibitem{Bernard:1995dp}
V.~Bernard, N.~Kaiser, and U.-G. Meissner,
\newblock Int.J.Mod.Phys. {\bf E4}, 193 (1995), hep-ph/9501384.

\bibitem{Becher:1999he}
T.~Becher and H.~Leutwyler,
\newblock Eur.Phys.J. {\bf C9}, 643 (1999), hep-ph/9901384.

\bibitem{Beringer:1900zz}
Particle Data Group, J.~Beringer {\em et~al.},
\newblock Phys.Rev. {\bf D86}, 010001 (2012).

\bibitem{Miller:2002ig}
G.~A. Miller,
\newblock Phys.Rev. {\bf C66}, 032201 (2002), nucl-th/0207007.

\bibitem{Blok:2008jy}
Jefferson Lab, H.~Blok {\em et~al.},
\newblock Phys.Rev. {\bf C78}, 045202 (2008), 0809.3161.

\bibitem{Huber:2008id}
Jefferson Lab, G.~Huber {\em et~al.},
\newblock Phys.Rev. {\bf C78}, 045203 (2008), 0809.3052.

\bibitem{Stoks:1992ja}
V.~G. Stoks, R.~Timmermans, and J.~de~Swart,
\newblock Phys.Rev. {\bf C47}, 512 (1993), nucl-th/9211007.

\bibitem{Budd:2003wb}
H.~S. Budd, A.~Bodek, and J.~Arrington,
\newblock (2003), hep-ex/0308005.

\bibitem{Koch:1980ay}
R.~Koch and E.~Pietarinen,
\newblock Nucl.Phys. {\bf A336}, 331 (1980).

\bibitem{Bernard:2007zu}
V.~Bernard,
\newblock Prog.Part.Nucl.Phys. {\bf 60}, 82 (2008), 0706.0312.

\bibitem{Bernard:1996gq}
V.~Bernard, N.~Kaiser, and U.-G. Meissner,
\newblock Nucl.Phys. {\bf A615}, 483 (1997), hep-ph/9611253.

\bibitem{Kubis:2000zd}
B.~Kubis and U.-G. Meissner,
\newblock Nucl.Phys. {\bf A679}, 698 (2001), hep-ph/0007056.

\bibitem{Kucukarslan:2006wk}
A.~Kucukarslan and U.-G. Meissner,
\newblock Mod.Phys.Lett. {\bf A21}, 1423 (2006), hep-ph/0603061.

\bibitem{Coon:1987kt}
S.~Coon and R.~Barrett,
\newblock Phys.Rev. {\bf C36}, 2189 (1987).

\bibitem{Belushkin:2006qa}
M.~Belushkin, H.-W. Hammer, and U.-G. Meissner,
\newblock Phys.Rev. {\bf C75}, 035202 (2007), hep-ph/0608337.

\bibitem{Hammer:2003ai}
H.~Hammer and U.-G. Meissner,
\newblock Eur.Phys.J. {\bf A20}, 469 (2004), hep-ph/0312081.

\bibitem{Mergell:1995bf}
P.~Mergell, U.~G. Meissner, and D.~Drechsel,
\newblock Nucl.Phys. {\bf A596}, 367 (1996), hep-ph/9506375.

\bibitem{Nagels:1977ze}
M.~Nagels, T.~Rijken, and J.~de~Swart,
\newblock Phys.Rev. {\bf D17}, 768 (1978).

\bibitem{Gardestig:2004hs}
A.~Gardestig {\em et~al.},
\newblock Phys.Rev. {\bf C69}, 044606 (2004), nucl-th/0402021.

\bibitem{Nogga:2006cp}
A.~Nogga {\em et~al.},
\newblock Phys.Lett. {\bf B639}, 465 (2006), nucl-th/0602003.

\end{thebibliography}

\end{document}